\documentclass[prd,twocolumn,amsmath,amssymb,showpacs,floatfix,nofootinbib,preprintnumbers,superscriptaddress]{revtex4-1}
\usepackage[latin1]{inputenc}
\usepackage[normalem]{ulem}
\usepackage{amsmath}
\usepackage{amsfonts}
\usepackage{amssymb}
\usepackage{graphicx}
\usepackage{dcolumn}
\usepackage{subcaption}
\usepackage{bm}
\usepackage{ulem}
\usepackage{latexsym}
\usepackage{epsfig}
\usepackage{graphicx}
\usepackage[colorlinks=true]{hyperref}
\usepackage[dvipsnames]{xcolor}
\begin{document}
\title{Fundamental physics using the temporal gravitational wave background}
\author{Suvodip Mukherjee}\email{ s.mukherjee@uva.nl, mukherje@iap.fr}
\affiliation{Gravitation Astroparticle Physics Amsterdam (GRAPPA),
Anton Pannekoek Institute for Astronomy and Institute for High-Energy Physics,
University of Amsterdam, Science Park 904, 1090 GL Amsterdam, The Netherlands}
\affiliation{Institute Lorentz, Leiden University, PO Box 9506, Leiden 2300 RA, The Netherlands}
\affiliation{Delta Institute for Theoretical Physics, Science Park 904, 1090 GL Amsterdam, The Netherlands}
\author{Joseph Silk}\email{joseph.silk@physics.ox.ac.uk, silk@iap.fr}
\affiliation{Institut d'Astrophysique de Paris (IAP), UMR 7095, CNRS/UPMC Universit\'e Paris 6, Sorbonne Universit\'es, 98 bis boulevard Arago, F-75014 Paris, France}
\affiliation{The Johns Hopkins University, Department of Physics \& Astronomy, \\ Bloomberg Center for Physics and Astronomy, Room 366, 3400 N. Charles Street, Baltimore, MD 21218, USA}
\affiliation{Beecroft Institute for Cosmology and Particle Astrophysics, University of Oxford, Keble Road, Oxford OX1 3RH, UK}
\date{\today}

\begin{abstract}
We propose a novel probe  of fundamental physics  that involves the  exploration of
 temporal correlations between 
the multi-frequency electromagnetic (EM) signal and the sub-threshold GW signal or  stochastic gravitational wave background (SGWB) 
originating from coalescing binaries. This method will be useful for the detection of  EM counterparts associated with the sub-threshold/SGWB signal. 
Exploiting the time delay between concomitant emission of the gravitational wave  and  EM signals enables inference of  the redshifts of the contributing sources 
by studying the time delay dilation due to cosmological expansion, provided that the time-lag between the emission of gravitational wave signal and the EM signal acts like a standard clock. 
Measurement of the inevitable time-domain correlations between different frequencies of gravitational and EM waves, most notably in gamma-rays, will test several aspects of fundamental physics and gravitation theory, and enable a  new pathway for current and future gravitational wave telescopes to study the universal nature of binary compact objects to high redshifts. 

  \end{abstract}
\maketitle
\section{Introduction}
Multi-frequency observations of a coalescing binary object in both gravitational 
and electromagnetic (EM) waves were first made for the neutron star binary merger  GW170817 \cite{Monitor:2017mdv, PhysRevLett.119.161101, Abbott:2017xzu}. This event left its imprint in the gravitational wave (GW) signal, as well as on the EM signals from gamma rays to the radio domain \cite{GBM:2017lvd}. Such an event makes it possible to study both astrophysical and  cosmological aspects \cite{Abbott:2017xzu,Margutti:2017cjl, Cowperthwaite:2017dyu,PhysRevLett.120.172703}. Analogously to this relatively nearby event, there are multiple astrophysical sources (binary neutron stars (BNSs), neutron star-black holes (NS-BHs), and binary black holes (BBHs)) coalescing in the observational Universe at high redshifts that cannot be detected as individual events. Such sources can be classified into two groups: sub-threshold GW events and  astrophysical stochastic GW background (SGWB) \cite{Allen:1997ad,Phinney:2001di, Regimbau:2007ed,Zhu:2011bd, Mitra:2007mc, Thrane:2009fp, Talukder:2010yd,Romano:2016dpx,10.1093/mnras/stz3226,Ginat:2019aed,Nunes:2020rmr}.

EM counterparts from the GW sources are expected under several scenarios. For 
sources such as BNSs and NS-BHs, one expects to see EM counterparts from the jet ejecta and their shock interactions  with the interstellar medium (ISM) \cite{Li:1998bw,2012ApJ...746...48M,Barbieri:2019sjc, 2041-8205-752-1-L15}. For sources such as stellar origin BBHs detectable in the LIGO-Virgo frequency band \cite{TheLIGOScientific:2014jea, Martynov:2016fzi, PhysRevLett.123.231107, Acernese_2014,PhysRevLett.123.231108}, EM counterparts are expected if baryonic matter surrounds the coalescing black hole \cite{McKernan:2019hqs,PhysRevLett.124.251102}, notably including  neutrino-driven winds \cite{10.1111/j.1365-2966.2010.17600.x,Veres:2019hsd} in the presence of magnetic fields \cite{1977MNRAS.179..433B}, and/or if the BBHs are charged \cite{Zhang_2016}. For supermassive BBHs (SMBHs), detectable with the upcoming space-based GW detector LISA \cite{2017arXiv170200786A}, there are also likely to be EM counterparts due to the presence of surrounding baryonic matter.  \cite{2041-8205-752-1-L15, Haiman:2018brf,Palenzuela:2010nf,Farris:2014zjo,Gold:2014dta,Armitage:2002uu}. 

For detected GW sources, one can do a targeted sky search in the direction of the source, as was done for the GW sources GW170817 \cite{Monitor:2017mdv, PhysRevLett.119.161101, Abbott:2017xzu, 2017Sci...358.1556C, Soares-Santos:2017lru} and GW190521 \cite{PhysRevLett.124.251102, PhysRevLett.125.101102}. Such exploration allows us to validate or rule out theoretical models associated with the emission process and GW signals from the light curves obtained over multiple frequencies. Observation of such systems in the future will make it possible to understand the astrophysical properties of the sources by characterizing their light curves over a wide range of EM frequency channels\cite{2017Natur.551...64A, 2017Sci...358.1559K}. By using the anticipated population of such sources, we will be able to either validate existing theoretical models or be able to develop empirically developed models. While detected GW sources are going to be the guiding principle in the near future, they can only be observed in the relatively nearby Universe with the current generation GW detectors such as LIGO \cite{TheLIGOScientific:2014jea, Martynov:2016fzi, PhysRevLett.123.231107}, Virgo \cite{Acernese_2014,PhysRevLett.123.231108}, KAGRA \cite{PhysRevD.88.043007,Akutsu:2018axf}. There should be many sub-threshold GW sources contributing to the astrophysical SGWB, and these are equally likely to have  EM counterparts but are not detected as a counterpart to the GW sources.  {With aLIGO design sensitivity \cite{TheLIGOScientific:2014jea, Martynov:2016fzi, PhysRevLett.123.231107}, we can only detect individual BNS events up to about redshift $z\sim 0.1$. So, BNS sources at higher redshift contribute to the sub-threshold GW signal or SGWB signal. However, the EM telescopes operating over a broad range of frequencies also have access to the Universe above redshift $z=0.1$. {In particular, detection of gamma-ray emission from Fermi/Gamma- ray Burst Monitor (GBM) and Swift Observatory is possible beyond redshift $z=0.1$ \cite{Fermi-LAT:2013oro,Connaughton:2013dna,Patricelli:2016bkt,RodrigoSacahui:2019qda,Troja:2019ccb} and after the beginning of the observation from Vera Rubin Observatory \cite{2009arXiv0912.0201L}, and Roman Telescope \cite{Gehrels_2015, 2013arXiv1305.5425S}, it will be possible to detect kilonova using low-energy EM frequency channels also beyond redshift $z=0.1$ \cite{Chase:2021ood}.} As a result, while some of the GW sources cannot be detected as individual events, their EM counterparts (if they exist) can be detected using EM telescopes. So, if there is a way to connect those well-detected EM signals with the sub-threshold GW  or SGWB signals, then we can identify EM counterparts to these sources, and can also confirm the astrophysical origin of the sub-threshold GW signal or the SGWB signal.} Detection of the EM counterparts associated with such sources should allow us to study the redshift evolution of the astrophysical sources to high cosmological redshift. 

 {In this paper, we are proposing a novel idea to enable one to identify EM counterparts with associated sub-threshold GW signal or SGWB signals. The usual practice is to search for the EM counterpart to the well-detected GW sources by sending triggers of the GW events to multiple telescopes. What we are suggesting here is to perform the opposite approach for the sub-threshold or SGWB signals. We propose to search for GW signals in the time-ordered GW data at those times when there are transient EM sources detected at a particular sky direction in multiple frequency channels. The detected EM signals can be used as a guiding principle to classify the astrophysical GW signal and distinguish it from detector noise. We propose a technique of time-domain cross-correlation between the EM signal and the GW signal to identify the EM counterparts of sub-threshold GW signal and SGWB signals. We develop the formalism of time-domain cross-correlations and discuss how it can be used for not-well detected GW events to search for the EM counterparts. The time-domain cross-correlation between the transient EM signals with the GW signals will exhibit a strong correlation only when the signal is present in both GW and EM data. Otherwise, it will be consistent with zero. As a result, if there is a common origin of EM signals at multiple frequency bands, and there is also a corresponding signal in GW data, then it can be distinguished from the GW detector noise or any other non-common origin of the signal. }  {The method proposed here is also applicable for the search for possible correlations with neutrino signals such as GW-neutrino searches and EM-neutrino searches. In this work, we introduce the basic formalism of the time-domain correlation technique for the search of EM counterparts to the sub-threshold GW events. More detailed studies for individual missions operating at specific EM frequency channels are required to understand the application of the time-domain cross-correlation technique which will depend on the mission-specific transient follow-up strategies. These will be explored in a series of works in the future.}

\section{Basic Framework}
The energy density of the SGWB with respect to the critical energy density of the Universe $\rho_c c^2= 3H_0^2c^2/8\pi G$ can be expressed in terms of the number of compact objects $n(z, t_r(z), \theta, \hat \alpha)$ coalescing in direction $\hat \alpha$ and emitting a GW signal per comoving volume between time $t_r(z)$ to $t_r(z+\Delta z)$ for the astrophysical parameters denoted by $\theta \in$ \{mass of the coalescing binaries, spin, inclination angle\} with probability $p(\theta)$, as
\cite{Allen:1997ad,Phinney:2001di, Regimbau:2007ed,Zhu:2011bd, Mitra:2007mc, Thrane:2009fp, Talukder:2010yd,Romano:2016dpx}
\begin{align}\label{basic-1}
    \begin{split}
        \Omega_{GW} (f, t, \hat \alpha)=& \frac{1}{\rho_cc^2}\int dz \int d\theta p(\theta) \frac{n(z, t_r(z), \theta,\alpha)}{(1+z)} \\& \times \frac{dE_{GW}}{d\ln f_r}(f_r, \theta, t_r(z), \hat \alpha)\bigg|_{f_r= (1+z)f}, 
    \end{split}
\end{align}
where $t_r(z)$ is the time in the source frame and $t$ is the time in the observer's frame, and the observed frequency  is related to the source frequency by the relation $f= f_r/(1+z)$. The compact objects contribute to the SGWB signal in its inspiral, merger, and ring-down phase. The  corresponding energy spectrum per logarithmic frequency bin is \begin{align}\label{sgwb-1a}
    \begin{split}
        \frac{dE_{GW}(\theta)}{d f_r}= \frac{(G\pi)^{2/3}\mathcal{M}_c^{5/3}}{3} \mathcal{G}(f_r), 
    \end{split}
\end{align}
where $\mathcal{G}(f_r)$ captures the frequency dependence of the GW signal which can be modeled as  \cite{Ajith:2007kx}
\begin{equation}\label{fr-dep-1}
    \begin{split}
        \mathcal{G}(f_r)=
        \begin{cases}
        f_{r}^{-1/3} \, \text{for}\, f_r < f_{merg},\\  
         \frac{f_{r}^{2/3}}{f_{merg}}\, \text{for} \, f_{merg} \leq f_r < f_{ring},\\
         \frac{1}{f_{merg}f^{4/3}_{ring}}\bigg(\frac{f_{r}}{1+(\frac{f_r-f_{ring}}{f_w/2})^2}\bigg)^2\, \text{for}\,  f_{ring} \leq f_r < f_{cut},
         \end{cases}
    \end{split}
\end{equation}
where $f_{x}= c^3(a_1\eta^2 + a_2\eta +a_3)/\pi G M$ in terms of total mass $M= m_1 + m_2$ and symmetric mass ratio $\eta= m_1m_2/M^2$. The values of $a_1, a_2$, and $a_3$ for different $f_x$ are mentioned in table \ref{tab:params} \cite{Ajith:2007kx}. For   given coalescing binaries of masses $m_1$ and $m_2$, the binaries will be emitting GWs in the inspiral part up to frequency $f_{merg}$, followed by the ringdown part up to frequency $f_{ring}$, and will stop  emitting GW signals after $f_{cut}$. $f_w$ denotes the width of the Lorentzian function \cite{Ajith:2007kx}. The frequency of the emitted GW signal is inversely proportional to the total mass $M$ of the coalescing binaries as shown in Eq. \eqref{fr-dep-1}, resulting in a higher observed frequency from the ring down phase for lighter total masses than for the systems composed with heavier total mass. 

\begin{table}
    \centering
\begin{tabular}{|p{1.5cm}|p{1.5cm}|p{1.5cm}|p{1.5cm}|}
\hline
         $f_i$ & $a_1$ ($\times 10^{-1}$) &$a_2$ ($\times 10^{-2}$) &$a_3$ ($\times 10^{-2}$)\\
         \hline
         \hline
           $f_{merg}$ & $2.9740$ & $4.4810 $ & $9.5560$\\
          $f_{ring}$ & $5.9411$ & $8.9794$ & $19.111$\\
          $f_{cut}$ & $8.4845$ & $12.848$ & $27.299$\\
          $f_{w}$& $5.0801$ & $7.7515$ & $2.2369$\\
          \hline
    \end{tabular}
    \caption{We show the values of the parameters to obtain the frequency $f_{merg}$, $f_{ring}$, $f_{cut}$, and $f_w$ denoted by the functional form $f_i=c^3 (a_1\eta^2 + a_2\eta +a_3)/\pi G M$ \cite{Ajith:2007kx}.}
    \label{tab:params}
\end{table}

The number of overlapping coalescing binaries which contribute to the SGWB signal over a particular time $t$ at a frequency $f$ can be determined using the duty cycle. The duty cycle for the GW sources contributing to the SGWB signal is defined as \cite{Wu:2011ac, Rosado:2011kv}  
\begin{align}\label{duty-cyc}
    \begin{split}
\frac{d\mathcal{D}}{df}= \int dz \, \dot n(z) \frac{d\tau_d}{df},    \end{split}
\end{align}
where $\dot n (z)$ is the global event rate as a function of the cosmological redshift $z$ and the duration that the GW signal spends at frequency $f$ is 
\begin{eqnarray}\label{tauf}
\frac{d\tau_d}{df}= \frac{5c^5}{96\pi^{8/3} G^{5/3} \mathcal{M}_z^{5/3} f^{11/3}},
\end{eqnarray}
where, $\mathcal{M}_z$ is the redshifted chirp mass of the GW sources. For values of the duty cycle $\frac{d\mathcal{D}}{df} >1$, the observed SGWB is dominated by the overlapping GW sources. In the opposite limit, the SGWB is going to be sporadic. 

We show the duty cycle in Fig. \ref{fig:duty} for a  {constant} merger rate $\dot n_{BBH}= 20\, \rm Gpc^{-3}\rm yr^{-1}$, $\dot n_{NS-BH}= 30\, \rm Gpc^{-3}\rm yr^{-1}$, and $\dot n_{BNS}= 300\, \rm Gpc^{-3}\rm yr^{-1}$ \cite{LIGOScientific:2018mvr,Abbott:2020uma,Abbott:2020gyp}.  {A smaller/higher value of the merger rate will reduce/increase the duty cycle of the GW signal.}  The duty cycle {for stellar origin compact objects} is going to be less than unity for SGWB frequency $f>20$ Hz, and as a result, the sources contributing to the SGWB are non-overlapping  {and can be distinguished}. {The differences in the duty cycle are going to show different temporal behaviors of the SGWB signal for different kinds of compact objects (such as BNSs, NS-BHs, and BBHs), that can be used to differentiate between these sources \cite{10.1093/mnras/stz3226}.  {The merger rate of GW sources at high redshift is not yet known. The temporal behavior of the SGWB will be useful for determining the high redshift merger rate of the GW sources \cite{10.1093/mnras/stz3226}.}   {The current upper bounds on the all sky-integrated strength of SGWB and directional SGWB from the O1+O2+O3 data of LIGO-Hanford, LIGO-Livingston, and Virgo is  $3.4 \times 10^{-9}$ at the reference frequency $f=25$ Hz \cite{LIGOScientific:2019vic, Abbott:2021xxi} and $(0.56-9.7) \times 10^{-9}$ sr$^{-1}$  \cite{LIGOScientific:2019gaw, LIGOScientific:2021zam} respectively for $\Omega_{GW} (f) \propto f^{2/3}$.} The strength of the SGWB  power spectrum is a direct probe of the merger rate of high redshift sources \cite{10.1093/mnras/stz3226, Boco:2019teq, Callister:2020arv} and can also be used for estimating the event rate of lensed systems \cite{Mukherjee:2020tvr,PhysRevLett.125.141102}. 

 {For sub-threshold GW events or loud GW events, we can write the observed GW flux $\mathcal{F}^{GW} (\hat \alpha, t)$ at a  sky direction $\hat \alpha$, and at a time $t$ as}
\begin{align}\label{non-sgwb}
    \begin{split}
  \mathcal{F}_{GW} (\hat \alpha, t)= \frac{c^3}{16\pi G}(|h_+(\hat \alpha, t)|^2 + |h_\times(\hat \alpha, t)|^2)
\end{split}
\end{align}
 {here, $h_{+,\times}(\hat \alpha, t)$ denotes the observed GW strain at time $t$ in the sky direction $\hat \alpha$. The observed strain is assumed to be detected with matched-filtering signal-to-noise ratio (SNR) above eight, or sub-threshold events with SNR below eight but distinguishable from the SGWB. In this expression we have assumed that  each signal is isolated and non-overlapping. For overlapping signals, one can write the expression in terms of the sum over multiple events along a particular sky direction.}   

\begin{figure}
    \centering
    \includegraphics[trim={0cm 0.cm 1.cm 0.5cm},clip,width=0.5\textwidth]{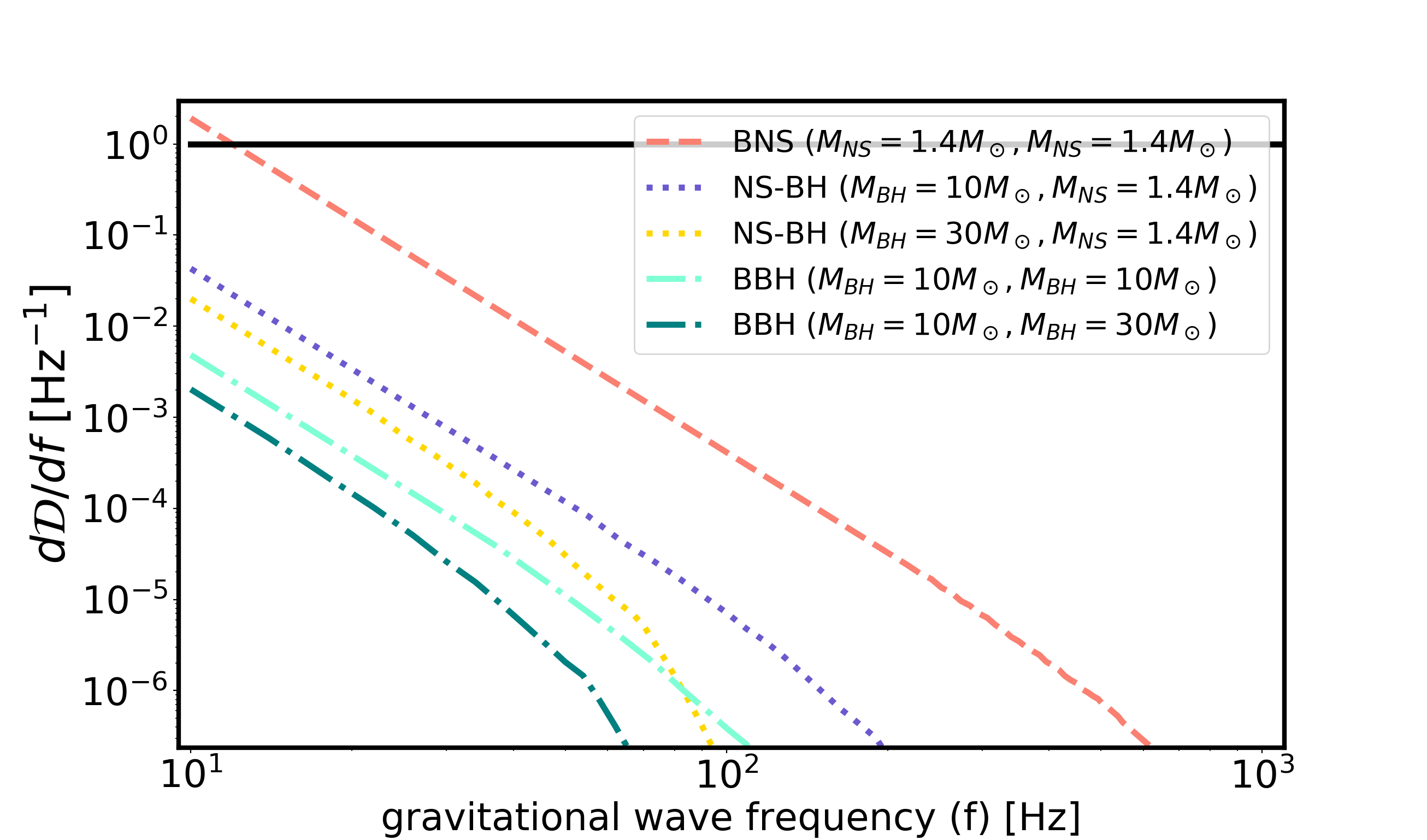}
    \captionsetup{singlelinecheck=on,justification=raggedright}
    \caption{{We show the duty cycle as a function of frequency for the BNS, NS-BHs and BBHs  in the frequency range observable from ground-based GW detectors. For  non-overlapping GW sources, the values of  the duty cycle are less than unity.}}
    \label{fig:duty}
\end{figure}

\begin{figure}
    \centering
    \includegraphics[trim={0cm 0.cm 0.cm 0.cm},clip,width=.5\textwidth]{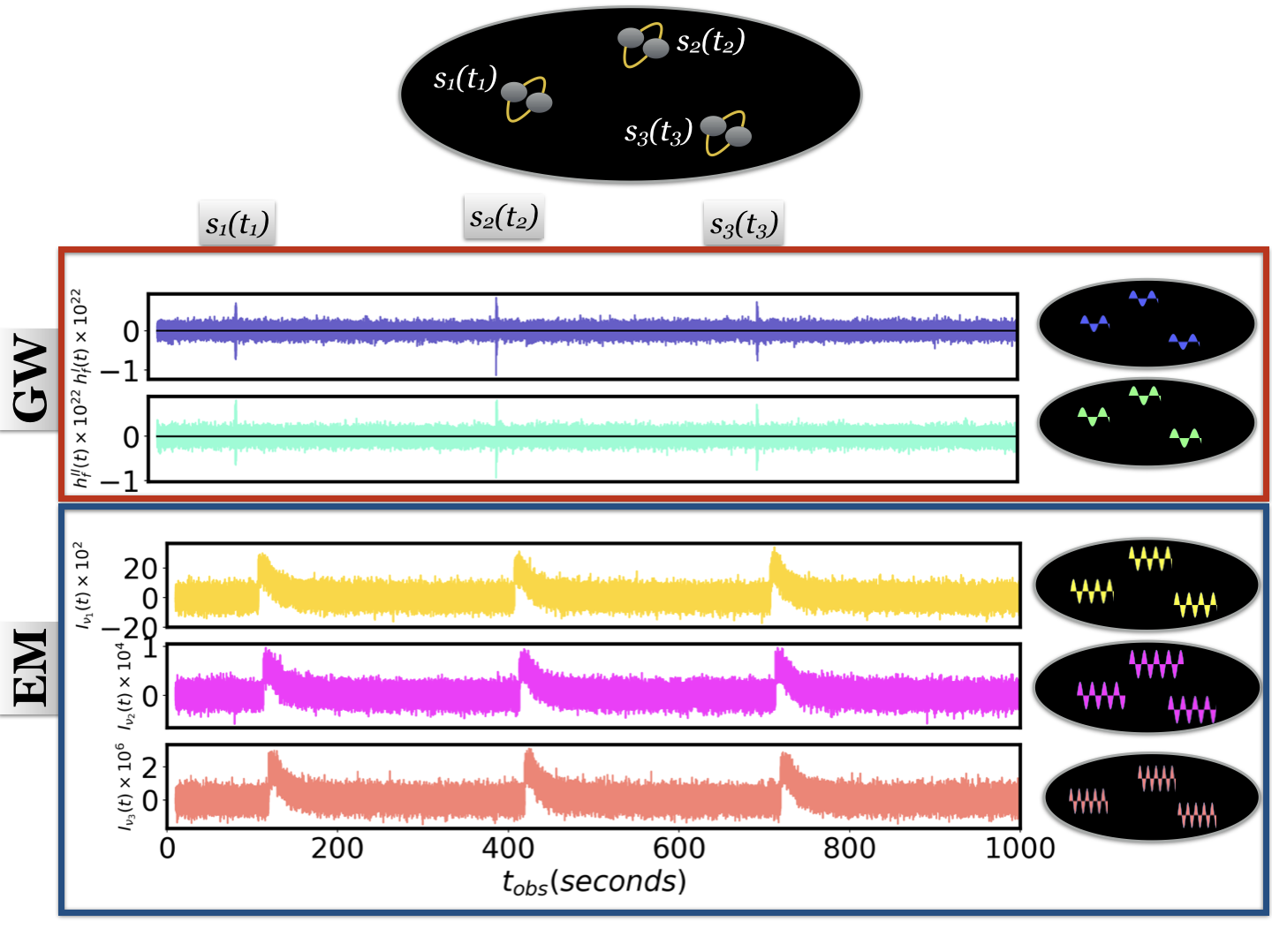}
    \captionsetup{singlelinecheck=on,justification=raggedright}
    \caption{{A schematic diagram showing the basic principle behind the time-domain correlation between SGWB and EM signals from the same sky location. In the upper panel, we show a cartoon data at a GW frequency $f$ for two different detector setups denoted by $I$ and $II$. The three (exaggerated) GW signals $s(t_i)$ are shown at time $t_i$ for the purpose of illustration, assuming three different sky locations shown by the map on the right. In the bottom panel, we show the cartoon time-series data of the EM signals $s(t_i +\Delta t)$ from these sources after time $\Delta t$ at three different EM frequencies $\nu_1, \nu_2, \nu_3$. The corresponding sky map of the EM signal is shown on the right for three different frequency channels. For the purpose of  illustration, we have considered a simple case of emission of the EM signal after the same time difference $\Delta t$ in all the frequency channels.}}
    \label{fig:schematic}
\end{figure}

Similarly, the intensity of the EM signal $\mathcal{I}_\nu (t, \hat \alpha)$ from all the sources emitting along the sky direction $\hat \alpha$ can be expressed as 
\begin{align}\label{basic-2}
    \begin{split}
        \mathcal{I}_\nu (t, \hat \alpha)=& \iint \frac{c dz d\theta p(\theta)n(z, t_r(z), \theta, \alpha)}{4\pi H(z)(1+z)^3} \\ & \times L_{EM}(\nu_r, \theta, t_r(z), z, \hat \alpha),
            \end{split}
\end{align}
where $L_{EM}(\nu_r, \theta, t_r(z), z, \hat \alpha)$ is the luminosity of the EM signal at the source frequency $\nu_r= (1+z
)\nu$ from a binary system with intrinsic source parameters  {denoted by} $\theta \in$ \{mass of the coalescing binaries, spin, inclination angle\} at the time in the source frame $t_r(z)$ from sky direction $\hat \alpha$. The luminosity of the EM signal $L_{EM} (\nu_r)$ depends on the properties of the astrophysical systems and on the stage of the merger, but the true nature of $L_{EM} (\nu)$ for different sources is not yet known from observations.

 {In the remainder of the paper, we will mainly discuss the time-domain cross-correlation between the SGWB signal and the EM signal. However, this formalism can be also directly applied to sub-threshold events or well-detected events.}

\section{Method} {The emission of the EM signals at frequency $\nu$ from GW sources emitting at frequency $f$ (during inspiral, or merger, or ringdown phase) will lead to an inevitable correlation in the time domain with a time-lag $\Delta t_{f\nu}= t_f - t_\nu$ at fixed sky direction.} A schematic diagram is shown in Fig. \ref{fig:schematic}  explaining the underlying principle of the temporal correlation. GW sources contributing to the SGWB can be detected in two different detectors denoted as $h_I$ and $h_{II}$ at an observed frequency $f$ (shown in the upper two panels). If the binary source emitting the GW signal also emits an EM signal at different frequencies $\nu_1, \nu_2$, and  $\nu_3$ (three lower panels) after a time delay $\Delta t_{f\nu_i}$, then it will appear in the same sky location. 

The expected time-domain correlation function can be written as
\begin{align}\label{ther-1}
    \begin{split}
        C_{f\nu}(t_{obs}, &\Delta t_{f\nu}, \hat \alpha)= \iint \frac{dzd\theta\, p(\theta) n^2(z, t_{f_r}(z),\theta)}{4\pi \rho_c c H(z)(1+z)^4}  \\ & \times  \frac{dE_{GW}}{d\ln f_r}(f_r, \theta, t_{f_r}(z), \hat \alpha)L_{EM}(\nu_r, \theta, t_{\nu_r}(z),\hat \alpha) \\ & \times  \delta (t_{f_r}(z)-t_{\nu_r}(z) -\Delta t_{f_r\nu_r}(z)),
    \end{split}
\end{align}
where, $\Delta t_{f_r\nu_r}= \Delta t_{f\nu}/(1+z)$ is the relation between the correlation time-scale between the source frame and the observer frame, and $t_{f_r}$ (and $t_{\nu_r}$) are the time in the source frame for the SGWB signal (and EM signal).  {If a GW source contributing to the SGWB has  
follow-up EM emission after a  time delay $\Delta t_{f_r\nu_r}$ after the end of the inspiral stage of the binary in its rest frame, then} assuming the general theory of relativity, the observed time delay  is 
\begin{align}\label{reds-1}
    \begin{split}
        \Delta t_{f_{obs}\nu}=&  \int_{f_{obs}}^{f_{merg}} \frac{d\tau}{df}df  + (1+z)\Delta t_{f_r\nu_r},
    \end{split}
\end{align}
where the first term denotes the duration of the GW signal between the observed frequency $f_{obs}$ and the frequency at the end of the inspiral phase $f_{merg}$ in the observer's frame, after which the EM emission at frequency $\nu_r$ can be characterised by the time duration $\Delta t_{f_r\nu_r}$.  Since from the SGWB, we cannot measure the phase of the GW signal, we cannot infer the stage of the inspiraling binary system which we observe at frequency $f_{obs}$. As  a result, the first term is an unknown shift for every pair of combinations of $f_r$ and $\nu_r$. However as this time-shift is constant for multiple frequency bands of the EM signal, we can exploit the characteristic time-lag between GW signal and EM signal at two different frequencies $\nu$ and $\nu'$ for the same SGWB signal by  
\begin{align}\label{reds-2}
    \begin{split}
        \Delta t_{f_{obs}\nu\nu'} \equiv \Delta t_{f_{obs}\nu}- \Delta t_{f_{obs}\nu'}=& (1+z)\Delta t_{f_{r}\nu_r\nu_{r}'},
    \end{split}
\end{align}
where $\Delta t_{f_{r}\nu_r\nu_{r}'}\equiv \Delta t_{f_r\nu_r}- \Delta t_{f_r\nu'_r}$ is the difference between the characteristic time-scale for GW emission at  frequency $f_r$ and EM emission at frequencies $\nu_r$ and $\nu_r'$ in the source rest frame. Depending on whether the characteristic time-scale in known or not, one can use the above equation to find the redshift.

 {For the signals which are detected as individual loud events or sub-threshold events, we can write the time-domain cross-correlation as}
\begin{align}\label{ther-2}
    \begin{split}
        C_{f\nu}(t_{obs}, &\Delta t_{f\nu}, \hat \alpha)= \iint \frac{dzd\theta\, p(\theta) n(z, t_{f_r}(z),\theta)}{4\pi H(z)(1+z)^3}  \\ & \times  \mathcal{F}_{GW}(f_r, \theta, t_{f_r}(z), \hat \alpha)L_{EM}(\nu_r, \theta, t_{\nu_r}(z),\hat \alpha) \\ & \times  \delta (t_{f_r}(z)-t_{\nu_r}(z) -\Delta t_{f_r\nu_r}(z)),
    \end{split}
\end{align}
 {where $\mathcal{F}_{GW}(f_r, \theta, t_{f_r}(z), \hat \alpha)$ is the Fourier transform of the GW loud/sub-threshold events mentioned in Eq. \eqref{non-sgwb}. For a non-overlapping GW signal, the theoretical correlation function is only the product of the EM signal and the GW signal within the time-window when both the signals are present in both  data sets. In the remaining of the analysis we will only discuss  the SGWB time-domain correlation. But our results will also be applicable to the individual loud/sub-threshold events. We will explore in detail the time-domain correlation for the individual events in  future work.}  

\textit{For a fixed characteristic time-scale: } 
If the characteristic time-scale $\Delta t_{f_r\nu_r\nu_r'}$ is driven primarily by the scales associated with the astrophysical compact objects, then it is unlikely to depend on the cosmic epoch and its source redshift $z$.  {As a result, we can expect the characteristic time-scale to be unique and act like a \textit{standard clock} for similar kinds of binary systems.  {The time difference $\Delta t_{f_{r}\nu_r\nu_{r}'}$ in the rest frame of the source is going to vary for different types of binaries (BNS, NS-BH, BBHs, SMBHs), but can be characterised as standard clock for each individual type of binaries. By using the EM spectrum $\mathcal{I}_\nu$, we can characterize the type of sources contributing to the SGWB and the characteristic time-scale $\Delta t_{f_{r}\nu_r\nu_{r}'}$ can be modelled from the low redshift individual events.}} Using such a standard clock, we can infer the source redshift using Eq. \eqref{reds-2}. The corresponding minimum variance estimator for inferring the redshifts of the GW sources by combining all $N$ frequency channels is
\begin{align}\label{basic-4}
    \begin{split}
     1+ \hat z=& \frac{\sum_{N} \bigg(\frac{1}{2\sigma^2_{t_{GW}} (f)+ \sigma^2_{t_{EM}}(\nu) +  \sigma^2_{t_{EM}}(\nu')}\frac{\Delta t_{f\nu\nu'}}{\Delta t_{f_r\nu_r\nu'_r}}\bigg)}{\sum_{N} \frac{1}{2\sigma^2_{t_{GW}} (f)+ \sigma^2_{t_{EM}}(\nu) +  \sigma^2_{t_{EM}}(\nu')}}
    \end{split}
\end{align}
where $\sigma^2_{t_{GW}} (f)$ and $\sigma^2_{t_{EM}} (\nu)$ are the measurement errors associated with the arrival times of the GW and EM wave. The variance $\sigma^2_z$ on inferred redshift $\hat z$ is $\sigma^2_z= \bigg(\sum_{N} \frac{1}{2\sigma^2_{t_{GW}} (f)+ \sigma^2_{t_{EM}}(\nu) +  \sigma^2_{t_{EM}}(\nu')}\bigg)^{-1}$. The time difference mentioned above is measured using the time-difference between the correlation time between the EM signal at frequency $\nu$ (or $\nu'$) and the GW signal at frequency $f$. As a result, it depends on both the time accuracy of the GW signal and the EM signal.  {For  a kHz sampling rate of the GW signal and $<5\%$ measurement accuracy of the timing measurement in the gamma-ray frequency band, we can measure the redshift by this avenue with an accuracy $\sigma_z \lesssim$ (few) $\times 10^{-2}$ \footnote{We assume a $<5\%$ error on the measurement of the time delay in the gamma-ray frequency band following the error-bar on the measurement of the time-delay made for GW170817 \cite{GBM:2017lvd}.}} from a single combination of the frequency channels $f$, $\nu$ and $\nu'$. A further reduction in the error bars (by $1/\sqrt{N}$) is possible by combining all the channels.  {An alternative way to find the redshifts of the SGWB sources can be through identifying the host galaxy from a photometric/spectroscopic follow-up using telescopes \cite{2009arXiv0912.0201L, 2012arXiv1208.4012G, 2013arXiv1305.5425S, Aghamousa:2016zmz,2010arXiv1001.0061R, Stratta:2017bwq, Dore:2018kgp, Dore:2018smn, Bellm_2018},  after observing the EM counterparts for these signals,  predicted here to reveal  a strong time-domain correlation with the GW signal.  {The time delay between the GW signal and the EM signal will depend on the astrophysical systems and the structure of the astrophysical systems, as well as its environment. In such a scenario, the time delay will not be unique for every source, but will depend on the property of the host galaxy. In such a scenario measurement of the redshift using the time delay will be difficult. However, a measurement of the time delay cross-correlation signal is still be useful in learning the property of the astrophysical sources, its environment, and the dependence of the time-delay between the emission of the signal between the GW and EM signal at different frequency channels.}

\textit{For a random characteristic time-scale}: If the characteristic time scale is random and varies for each source, then one cannot use Eq. \eqref{reds-2} to obtain the redshift. In such a case, the redshift of the EM signal can be measured by host identification using  photometric or spectroscopic galaxy catalogs, available from the EM surveys \cite{2009arXiv0912.0201L, 2012arXiv1208.4012G, 2013arXiv1305.5425S, Aghamousa:2016zmz,2010arXiv1001.0061R, Stratta:2017bwq, Dore:2018kgp, Dore:2018smn, Bellm_2018}. The uncertainties in redshift $\sigma_z$ for photometric and spectroscopic galaxy catalog are expected to be of the order $\mathcal{O}(10^{-2})$ and $\mathcal{O}(10^{-4})$ respectively. However by estimating the redshift $\hat z$ from the host galaxy, one can  infer the characteristic time-scale $\Delta t_{f_{r}\nu_r\nu_{r}'}= \Delta t_{f_{r}\nu_r\nu_{r}}/(1+\hat z)$.

By estimation of the redshift either using the standard clock method (given in Eq. \ref{basic-4}) or by the host galaxy identification, we can write Eq. \eqref{ther-1} as a tomographic estimate at every redshift (see Eq. \eqref{reds-2}). After integrating over the observation time and sky directions, we can obtain the time-domain correlation signal as \footnote{The angular bracket $\langle.\rangle_{t_{obs}, \hat \alpha} \equiv \frac{1}{T_{obs}4\pi}\int d^2\hat \alpha\int dt$}
\begin{align}\label{ther-2}
    \begin{split}
        C_{f\nu}(\Delta t_{f\nu})= &\int \frac{d\theta\, p(\theta) n^2(z, t_{f_r},\theta)}{ \rho_c c H(z)(1+z)^4}   \bigg\langle \frac{dE_{GW}}{d\ln f_r}(f_r, \theta, t_{f_r}(z))\\ & \times L_{EM}(\nu_r, \theta, t_{f_r}(z)+ \Delta t_{f_r\nu_r}(z), z) \bigg\rangle_{t_{obs}, \hat \alpha}.
    \end{split}
\end{align}
 This signal depends on the intrinsic source property of the astrophysical system (the term present in the angular brackets), the number density of emitting systems $n(z, t',\theta)$, and the expansion history of the Universe $H(z)$. The spectral shape of the time-domain correlation signal depends on the spectral shape of the EM signal $L_{EM}$.  {Even in the absence of a characteristic time-scale $\Delta t_{f_r\nu_r}$, the time-domain cross-correlation signal between GWs and EM signals will exist and can be used to explore the properties of the astrophysical sources.} {
 For the correct association of the EM signal with the SGWB signal, we require that the observed characteristic time-scale $\Delta t_{f\nu}$ between  GW frequency $f$ and EM frequency $\nu$ is shorter than the time gap between the emission of two signals $\Delta t_{GW}$ at the same sky location in the SGWB map, i.e. $\Delta t_{f\nu}< \Delta t_{GW}$. $\Delta t_{GW}$ is related to the  event rate by the relation $\Delta t_{GW}= \bigg(\int \frac{dz \,c\, d^2_c\, \dot n(z)}{H(z)(1+z))}\bigg)^{-1}$, where $d_c$ denotes the comoving distance to redshift $z$. }
 
 Using the map of SGWB   $\Omega^{obs}_{GW}(f,t,\hat \alpha)= \frac{(32\pi^3f^3)}{3H_0^2}\delta (f-f') \langle h_I(f,t, \hat \alpha)h^*_J(f', t, \hat \alpha) \rangle$ \cite{Allen:1997ad} at frequency $f$ \footnote{Using short-time Fourier transformation of a time-ordered data $d(t')$ which is defined as $d(t, f)= \int_{t-\tau/2}^{t+\tau/2} d(t')e^{-2i\pi ft'} dt'$}, we can write  an estimator for the time-domain correlation with the intensity map $\mathcal{I}^{obs} (\nu,t,\hat \alpha)$ of the EM signal \footnote{$\mathcal{I}_\nu^{obs} (t,\hat \alpha)$ is the EM signal within the resolved sky resolution of the SGWB.}  as
\begin{widetext}
\begin{align}\label{basic-3}
    \begin{split}
        C_{f\nu}(t_{obs}, \Delta t_{f\nu}, \hat \alpha)&\equiv \bigg\langle   \bigg(\Omega^{obs}_{GW}(f,t',\hat \alpha)-\Omega^{b}_{GW}(f,t')\bigg)\bigg(\mathcal{{I}}^{obs}_\nu(t'+\Delta t_{f\nu}, \hat \alpha) - \mathcal{{I}}^{b}_\nu( t'+\Delta t_{f\nu})\bigg) \bigg\rangle,\\ &= \frac{1}{\delta t}\int^{t_{obs}+\delta t/2}_{t_{obs}-\delta t/2} dt' \bigg(\Omega^{obs}_{GW} (f, t', \hat \alpha) - \Omega_{GW}^{b} (f)\bigg)\bigg(\mathcal{{I}}^{obs}_\nu( t'+\Delta t_{f\nu}, \hat \alpha) - \mathcal{{I}}^{b}_\nu\bigg),
    \end{split}
\end{align}
\end{widetext}
where, $\delta t$ is the small time interval over which we average the signals between the EM signal at frequency $\nu$, with the GW signal at frequency $f$. The averaging time is much larger than the sampling time step $\delta t_{sam}$ of the SGWB signal, and small/similar to the temporal correlation scale $\Delta t_{f\nu}= (1+z)\Delta t_{f_r\nu_r}$, i.e. $\delta t_{sam} << \delta t \lesssim \Delta t_{f\nu}$. $\Omega_{GW}^{b}(f)$ and $\mathcal{I}^{b}_s(\nu)$ is the sky and time averaged background.   {If the observed intensity in some EM frequency bands are not sampled continuously, but the sky images are taken only after a discrete time interval $\Delta \tau_\nu$, then one replace the above integration into a discrete summation as} 
\begin{widetext}
 {\begin{align}\label{basic-3dis}
    \begin{split}
        C_{f\nu}(t_{obs}, \Delta t_{f\nu}, \hat \alpha)=\frac{1}{2n_{bin}} \sum^{t^{bin}_{obs}+n_{bin}}_{i=t^{bin}_{obs}-n_{bin}} \bigg(\Omega^{obs}_{GW} (f, t_i, \hat \alpha) - \Omega_{GW}^{b} (f)\bigg)\bigg(\mathcal{{I}}^{obs}_\nu( t_i+\Delta t_{f\nu, i}, \hat \alpha) - \mathcal{{I}}^{b}_\nu\bigg),
    \end{split}
\end{align}}
\end{widetext}
 {where $t^{bin}_{obs}$ is the bin which relates to the observed time $t_{obs}$, $t_i$ is the time-bin with index $i$, $\Delta t_{f\nu,i}= \Delta t_{f\nu}/\Delta \tau_\nu$ is the bin with which the signal should be correlated, and $n_{bin}= \delta t /2\Delta \tau_\nu$ denotes the number of bins over which the correlated signal is added over. For different telescopes and observation techniques, the value of $\Delta \tau_\nu$ is going to be different. Certainly, $\Delta \tau_\nu$ needs to be smaller than $\Delta t_{f\nu}$ to be able to capture the signal. For the prompt gamma-ray signal, the value of $\Delta \tau_\nu$ can be less than a second \cite{GBM:2017lvd, 2017Sci...358.1559K,Troja:2019ccb}, whereas for capturing the low-frequency EM signals, operating in X-ray, UV, optical, infrared, and radio, the value of $\Delta \tau_\nu$ can be of the order a few hours timescale \cite{GBM:2017lvd, 2017Sci...358.1559K,Cowperthwaite:2017dyu,Chase:2021ood}.
 So, by using either Eq. \eqref{basic-3} or Eq. \eqref{basic-3dis}, we can search for the possible correlated signal between GW and EM sectors. 

 In the absence of a correlated signal in both $\Omega_{GW}(f)$ and $\mathcal{I}^{obs}_s (\nu)$, the above estimator given in Eq. \eqref{basic-3} (or Eq. \eqref{basic-3dis}) vanishes. 
  {An excess intensity of the EM signal $\mathcal{I}_\nu^{obs} (t,\hat \alpha)$ over the background captures any additional transient signal. For different frequency channels of EM observations, we can construct the sky map of the difference in the intensity of the signal as a function of observation time, which can be cross-correlated with the SGWB signal.  
 The time-domain cross-correlation search between the GW signal and the EM signal needs to be carried out in multiple frequencies of both these sectors to separate an unrelated transient from an actual EM counterpart which is related to the EM signal.}  The cross-correlation technique in the time-domain makes it possible to isolate the signal from the noise  {and unrelated foreground signals and transients}, even when the signal is weak as the noise between different detectors is uncorrelated. This makes the time-domain cross-correlation a clean way to identify the EM counterparts of the SGWB sources.  {Using time-domain cross-correlations between the EM and GW signals, the EM signal can be used as a guide to distinguish between the true SGWB signal and noise.}  {The expression given in Eq. \eqref{basic-3} and Eq. \eqref{basic-3dis} is written in terms of the observed intensity difference in different frequency bands as a function of time. For the  EM frequency bands, we have written the above expression in terms of the difference in the intensity between two instants of time. The presence of an excess signal at a time $t$ over the sky background at the same sky direction is correlated with the excess signal in the GW  background signal. For an assured recovery of the EM counterpart to the GW signal and to reduce the confusion with the unrelated transient source, it is essential to have a measurement of the non-zero time-domain correlation for multiple EM frequency channels.  The total signal-to-noise ratio for the detection of the signal needs to be made by combining multiple frequency bands. We discuss more this in  Sec. \ref{snr}.}  {If the correlated signal between GW and EM is not present at different frequency channels, then it cannot be considered as EM counterpart related to the GW signal but should be considered as unrelated transient.}
 

\begin{figure}
    \centering
    \includegraphics[trim={1.cm 0.cm 1.cm 1.8cm},clip,width=.5\textwidth]{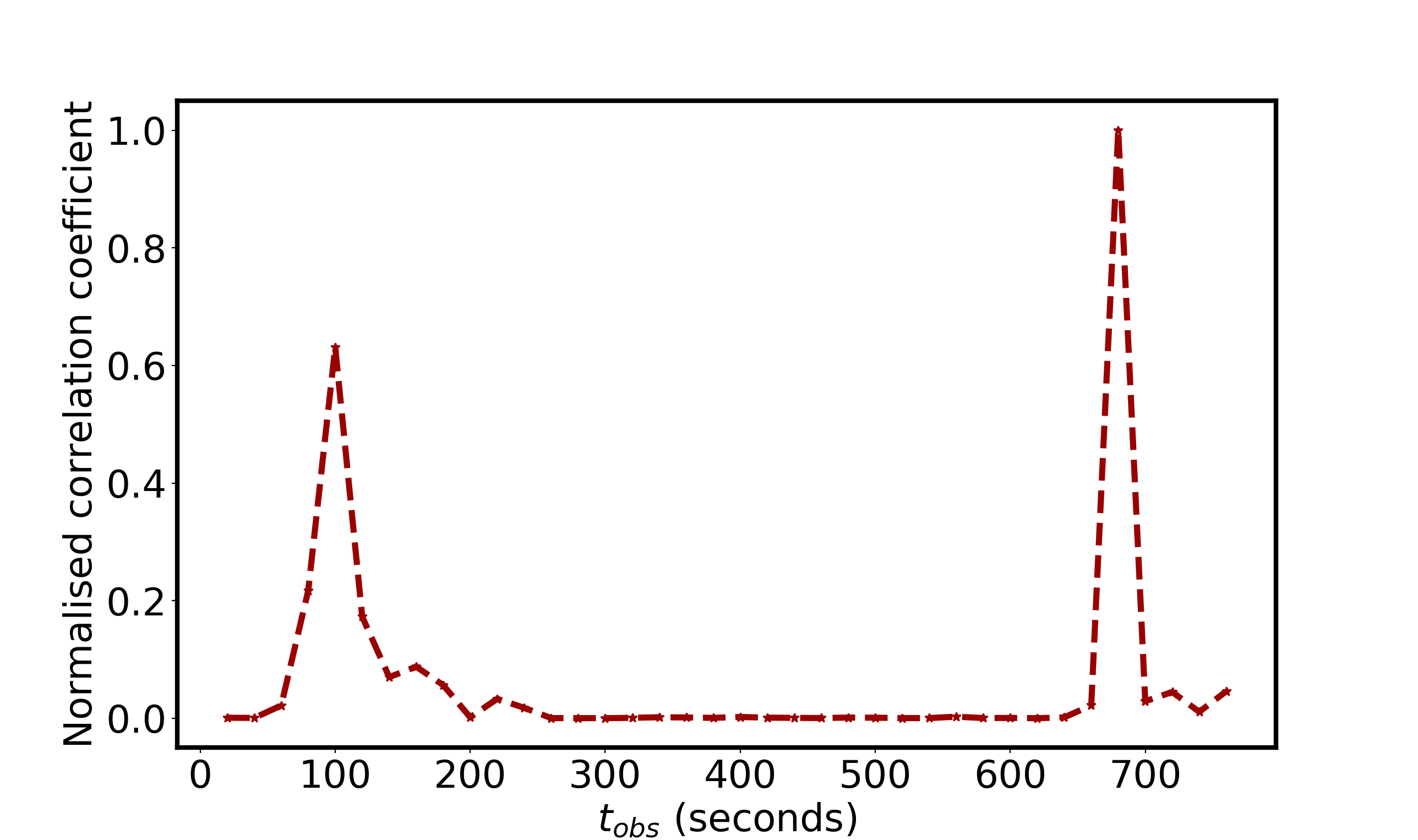}
    \captionsetup{singlelinecheck=on,justification=raggedright}
    \caption{{  {We show the normalized correlation coefficient between the mock time-series data of SGWB and EM signal as a function of the observation time $t_{obs}$ after performing $20$ second bin averages. The correlation between the two is large when the signals are present in both the data samples (SGWB and EM signal), otherwise, it does not show a strong correlation. The correlation time is the short amount of time during which the two signals show a correlation which is shorter than the observation time. The error on the correlation is larger than the amplitude of a single signal. So individual measurements are not statistically significant. One needs to combine several time-domain correlations to enhance the signal-to-noise ratio.  }}}\label{fig:correlation_time}
\end{figure}

\section{Application to mock samples} {To show a proof of principle of the time domain correlation, we implement this method on a simulated time-series data assuming two detectors ($h_I(t)$ and $h_J(t)$) and EM signals $\mathcal{I}_\nu(t)$ for two frequency channels. For the GW data, we obtain time-series data for a pair of GW detectors $i \in \{I, J\}$ as $h_i(t)= s(t)+ n_i(t)$, 
 {where $h_i(t)$ is the total GW strain in the $i^{th}$ detector, composed of the GW signal $s(t)$ and the Gaussian noise realizations $n_i(t, \hat x)$ for the $i^{th}$ detector with known noise power spectrum according to the design sensitivity of the LIGO-Virgo detectors \cite{Acernese_2014, Martynov:2016fzi}. The Gaussian noise realizations for different detectors are uncorrelated. For the GW signal, we consider binary neutron stars (BNSs) of individual mass $M_{NS}= 1.4\, M_\odot$ with two signals within a time-series data of length $1000$ seconds.}  For BNSs, with the currently known event rate from LVC \cite{LIGOScientific:2020kid}, the duty cycle is going to be less than one for a frequency range $f \geq 20$ Hz, which makes the sources non-overlapping in the time-domain.  So, we consider two non-overlapping sources contributing to the SGWB signal in a mock data duration of $1000$ seconds.  {The simulated time-series is assumed to have two injected sub-threshold GW signals at the level of one-fourth of the power spectrum of the instrument noise at frequency $f$ between $10$ Hz and $50$ Hz. The simulated signal is then added to the Gaussian noise having the power spectrum of aLIGO with different noise realizations for two different detectors.}  
 We apply the cross-correlation between the two simulated SGWB data sets  $h_I(t)$ and $h_J(t)$, to obtain the SGWB power spectrum  ($\hat \Omega^{obs}_{GW}(f, t, \hat \alpha)$) at frequency $f$  using \cite{Allen:1997ad,Mitra:2007mc, Thrane:2009fp, Talukder:2010yd,Romano:2016dpx}
 \begin{align}\label{sgwb-len-3}
    \begin{split}
     \Omega^{obs}_{IJ}(f, t, \hat \alpha) \equiv & \langle{h_I(f,t)h^*_J(f',t)}\rangle \\ =&\delta (f-f') \langle s_I(f,t, \hat \alpha)s^*_J(f', t, \hat \alpha) \rangle, 
    \end{split}
\end{align}

For the injected gravitational signals, we consider EM signals in the frequency bands $\nu_r=10^7$ Hz and $\nu_r= 10^{13}$ Hz, with a simple model of the light-curve written in terms of the time duration $\Delta t_{f_r\nu_r}$ after the merger of the GW sources and emission of the EM signal in the source rest frame by \cite{Beuermann:1999sj, vanEerten:2018vgj}
\begin{align}\label{emflux}
\begin{split}
    I_{\nu_r}(\Delta t_{f_r\nu_r})= A_\nu\bigg(\frac{\nu_r}{\nu_{0}}\bigg)^{-\beta}& \bigg(\bigg(\frac{\Delta t_{f_r\nu_r}}{t_p}\bigg)^{-\alpha_1 \kappa}\\& + \bigg(\frac{\Delta t_{f_r\nu_r}}{t_p}\bigg)^{\alpha_2 \kappa}\bigg)^{-1/\kappa},
\end{split}
\end{align}
where for this simple model, $t_p$ denotes the peak time whose value is taken as $9$ seconds, the rise and the decay parameters $\alpha_1$ and $\alpha_2$ are taken as $0.9$ and $2.0$ respectively. The smoothness parameter $\kappa= 2$ is considered in this analysis. The frequency dependence of the emission in considered as a power-law with the index $\beta=0.585$ with the pivot point $\nu_0= 10^7$ Hz. The amplitude of the flux $A_\nu$ is considered to be frequency-dependent with $A_{\nu= 10^7 Hz} = 10^{-2}$ Jy/sr and $A_{\nu=10^{13} Hz}= 10^{-3}$ Jy/sr. The standard deviation of the noise for the EM signal is considered to be constant in time as $\sigma_{I_\nu}/I_\nu= 0.5$ (which resembles only a $2\sigma$ detection of the EM signal).

We apply the estimator given in Eq. \eqref{basic-3} between the SGWB signal $\Omega^{obs}_{IJ}(f, t, \hat \alpha)$ with the intensity of the EM signal $\mathcal{I}_\nu (t +\Delta t)$ for different values of the time $\Delta t$ with the integration time scale $\delta \tau= 10$ seconds.  {The cross-correlation signal between the mock data samples shows a spike when the signal is present in both SGWB and EM signals.} The SGWB signal does not correlate with the noise in the EM signal and shows correlation only with the astrophysical source at the same spatial location separated by time difference time $\Delta t_{f\nu}$. {The time-domain cross-correlation between $\Omega_{GW}(f, t)$ and $\mathcal{I}_\nu (t)$ is shown in Fig. \ref{fig:correlation_time} after averaging over temporal window $10$ seconds using the estimator mentioned in Eq. \eqref{basic-3}. The correlation between the two is large only when the signal is present in both SGWB mock data  $\Omega_{GW}(f,t)$ and EM mock data $\mathcal{I}_\nu(t)$, otherwise the cross-correlation between the two time-series does not lead to a constructive signal. 

This is the key aspect that makes the time-domain correlation between the GWs and EM waves a potentially clean way to measure the EM counterparts to the signals which ordinarily are hidden in the noise of the GW detector. The fact that EM detector noise has zero mean, and is uncorrelated with the GW detector noise, makes it possible to find a strong correlation only when the signal is present in both the EM and SGWB data.}  {In Fig. \ref{fig:correlation_time} we show only the normalised mean cross-correlation signal.  {The normalized signal is defined as the correlation signal divided by the maximum value of the correlation signal. This is to show how different time-correlation signals will show up for different signal strengths.}
The cross-correlation indicates that if the signal is not present in both the GW and EM sectors, then the mean value of the cross-correlation vanishes. This is because if there is no excess signal over the background in the EM sector then the product of this with the GW background map vanishes.   {The inferred mean signal helps in identifying the possible counterpart and isolating unrelated transients, by combining multiple EM frequency channels.} The associated error on the measurement of the time-domain correlation will depend on the errors in both GW data and EM data, and we will discuss this in the next section. To make high signal-to-noise ratio measurements of the time-domain correlated signal, we need to do the cross-correlation study over time scales of years. If we assume that EM measurements have a higher signal-to-noise ratio than the SGWB, then the detection of the time-domain correlation signal improves as well. This is explained in more detail in the next section. }
\section{Signal-to-noise ratio for a network of GW detectors}\label{snr}
The signal-to-noise ratio (SNR) for the measurement of the correlation signal $C_{f\nu}(\Delta t_{f\nu})$ can be obtained by integrating over the observation time $T_{obs}$ and all-sky directions $\hat \alpha$. By estimating the SGWB signal over a frequency bandwidth $\Delta f$ centered at $f$, and estimating the EM signal over a frequency bandwidth $\Delta \nu$, centered at $\nu$, we can obtain the SNR as  
\begin{widetext}
 {\begin{align}\label{snr-1}
\begin{split}
    SNR (C_{f\nu}(\Delta t_{f\nu}))= \bigg(\sum_k\sum_{ij, i>j}\sum_{T_{bins}}\sum_{N_{pix}} \int_{f' - \Delta f'/2}^{f' + \Delta f'/2} df \int_{\nu' - \Delta \nu'/2}^{\nu' + \Delta \nu'/2} d\nu & C_{f\nu}(t_{obs}, \Delta t_{f\nu}, \hat \alpha)\mathcal{Z}^{-1}(\Delta\Omega^2_{N_{ij}}(f),\Delta \mathcal{I}^2_{N_k}(\nu), \alpha, \alpha')\\ & \times C_{f\nu}(t_{obs}, \Delta t_{f\nu}, \hat \alpha')\bigg)^{1/2},
    \end{split}
\end{align}}
\end{widetext}
 {where $\mathcal{Z}(\Delta\Omega^2_{N_{ij}}(f)\Delta \mathcal{I}^2_{N_k}(\nu), \alpha, \alpha')$ is the covariance matrix for the GW signal and EM signal between two sky positions $\alpha$ and $\alpha'$,} the summation over $k$ denotes all the EM emission detectable by  observing the sky at frequency $\nu$ with detector noise power spectrum $\Delta \mathcal{I}^2_{N_k} (\nu)$ (and assuming that the background intensity signal   $\mathcal{I}_s^b(\nu)$ is stationary).  {Here we do not assume the sky signal of the EM part to be Gaussian, but only assumes that the detector noise is Gaussian.} The sum over $\{ij\}$ denotes different pairs of GW detectors, $\Delta \Omega^2_{N_{ij}}(f)= 100\pi^4P_iP_jf^6/9H_0^4\gamma^2(f,t,\hat \alpha)$ is the SGWB noise power spectrum at frequency $f$ \cite{Allen:1997ad} which can be written in terms of the noise power spectrum of the $i^{th}$ GW detector $P_i$ (and $P_j$ for the $j^{th}$ detector), and the normalized overlap reduction function $\gamma(\hat \alpha, f, t)$ depends on the detector response function \cite{PhysRevD.48.2389, PhysRevD.46.5250}. $T_{bins}= T_{obs}/\delta t$ is the number of independent temporal bins available over the observation time $T_{obs}$, and $N_{p}$ is the number of independent sky patches in the overlapping observed sky area $f_{sky}(\nu)$ between the map of the SGWB and  the sky map of the EM signal at frequency $\nu$.  {The EM observations in every frequency band are not going to have full sky coverage. The overlapping sky coverage between the GW  and the EM observations are going to depend on the EM telescope and will be different for different EM frequency bands of the EM signal. In the next section (Sec. \ref{obs}), we briefly discuss a possible observational strategy in different frequency bands.}  The angular resolution of the SGWB map is diffraction-limited and given by $\Delta \Theta_{gw}= c/2fD$ where  $D$ is the distance between a pair of GW detectors. The above mentioned SNR improves with more observation time, more  GW detectors and EM detectors, better sky resolution, and larger fraction of observable sky area.

To simplify, above equation, if there are $N_{p}$ number of independent sky patches with the joint GW and EM signals detected all over the sky per year after combining all GW detectors, then we can write the signal-to-noise ratio (SNR) for the measurement of the $C_{f\nu}(\Delta t_{f\nu})$ as
\begin{equation}\label{snr-2}
\begin{split}
    SNR=   \bigg(\sum_{T_{obs}} \bigg(\frac{N_{p}T_{obs}}{ \text{yr}}\bigg) \frac{C^2_{f\nu}(\Delta t_{f\nu})}{\Delta\Omega^2_{N_{eff}}(f)\Delta \mathcal{I}^2_{N_{eff}}(\nu)}\bigg)^{1/2},
    \end{split}
\end{equation}
where $\Omega^2_{N_{eff}}(f)$ is the effective noise in the SGWB  from all pairs of detectors at GW frequencies $f$ within the bandwidth $\Delta f$. $\Delta \mathcal{I}^2_{N_{eff}}(\nu)$ is the effective noise in the intensity of the EM signal at frequency $\nu$ after combining all the detectors. {Although the high SNR SGWB measurement is going to be limited by the angular resolution $\Delta \Theta \sim 1$ radian \cite{10.1093/mnras/stz3226}, EM counterpart measurements should provide accurate identifications of sky localization. As a result,  angular scales which are unresolved in the SGWB signal can be probed with the aid of EM signals.  {Hence if we can pinpoint the sky locations using the multi-frequency EM observations, then we can combine all those sky positions to build up the total signal-to-noise ratio. As an example, suppose there are multiple GW signals with the same sky localization error, but separated by time. Then with only the GW sector, we are limited by the GW sky localization error. But using the EM observation, if we can identify sky positions of all these signal, then one combine all those sky directions.} As a result, the number of independent sky patches $N_p$ in the time-domain cross-correlation technique can be large and will be limited by the resolution of the instruments measuring the EM signal. 

\section{Observation strategy}\label{obs}
In the framework of time-domain correlations, we propose to cross-correlate the time-order EM data at different frequency bands with the time-ordered GW signal. EM radiation is expected to be emitted over a wide frequency range from gamma-ray to radio for timescales of a few seconds to years before/after the merger of the binaries. The transient nature of the EM counterparts, and the limited field of view of the EM telescopes in most of the frequency bands, make it challenging to follow up the EM counterparts to the GW sources. Moreover, due to the presence of a large number of EM emission signals which are not associated with the GW sources, it becomes even more challenging to identify the correct EM counterpart to the GW source. In the time-domain correlation proposed by us, both of these complications can potentially be overcome. We briefly describe the observation strategy in the following paragraphs.

We propose a three-step method to explore the association of the EM-GW signal. We propose to use (i) the time-ordered gamma-ray data to cross-correlate with the time-ordered GW data, to find whether a pair of EM-GW data signals are related, (ii) a follow-up search to measure the emission in other frequency bands of the EM signal in the direction of the sky for which the time-domain correlation signal between the gamma-ray and the GW data have shown non-zero signals  (iii) If the EM counterparts are detected also in other frequency bands, then to classify these signals as a possible joint-detection and perform a time-domain correlation between the GW and EM signals across the relevant EM frequency bands to understand their association by applying theoretical models or using the models developed empirically from the detected events \cite{2017Sci...358.1559K,Cowperthwaite:2017dyu, Dietrich:2020efo, Raaijmakers:2021uju}. In this setup, the EM data provides essential guidance to searching for the associated GW signal from the large volume of GW data. The reason for using the EM data as the trigger to search for the GW signal is due to the feasibility of EM telescopes to probe to higher redshifts than the currently ongoing GW detectors.  {The advantage of this proposed method of the time domain cross-correlation with multiple EM frequency bands makes it possible to distinguish between the unrelated transient sources. The method of time-domain cross-correlation at frequency bands only yields a non-zero effect when a signal is present in both GW and EM data. For a transient source, the signal will not show up at multiple EM frequency bands, in comparison to an actual associated signal event which will exhibit signatures in multiple EM frequency bands.}

Several previous studies have searched for EM counterparts for the GW sources \cite{GBM:2017lvd, 2017Sci...358.1559K,Cowperthwaite:2017dyu,Coughlin:2019xfb, Anand:2020eyg,GravityCollective:2021kyg} and the search for EM counterpart of the sub-threshold events are going to be even more difficult. With an upper bound on kilonova event rate $< 900$ Gpc$^{-3}$ yr$^{-1}$ \cite{Andreoni:2021ykx}, there will be a plethora of possible kilonova up to high redshift. The time-domain multi-frequency correlation technique will be useful to distinguish between an associated transient and unrelated transients. The time-domain cross-correlation study between the gamma-ray signal with the GW data can identify possible signals which can be associated with each other and will help in distinguishing between the unrelated transients and the transients which are related to GW sources and will allow to also reduce the sky localization to a few tens of square degrees to a few hundreds of square degrees \cite{Connaughton_2015,Connaughton:2016umz,Goldstein:2017mmi}. The follow-up search at a smaller sky volume will be possible from the lower frequency  channels such as X-ray, UV, optical, infrared, and radio to find the sources which exhibit time-domain correlation with the sub-threshold GW signal and gamma-ray signal. 
The main advantage of this technique is it can potentially open up the possibility of detecting more GW170817-like events and understand the true nature of compact objects. Furthermore, using the proposal of this work, weak GW signals can be isolated from the GW noise with the aid of EM data, which is currently not possible. 

 {The successful scientific outcome from the time-domain cross-correlation between the EM data and the GW data depends on the availability of EM telescopes in multiple frequency bands and the availability of networks of GW detectors. On the EM side, it is necessary to monitor a large fraction of the sky at different frequency bands with high cadence and to reach up to high cosmological redshifts. Upcoming missions such as the Vera Rubin Observatory \cite{2009arXiv0912.0201L}, Roman Telescope \cite{Gehrels_2015, 2013arXiv1305.5425S}, the Gravitational-wave Optical Transient Observer (GOTO) \cite{Dyer:2020dmo}, and also proposals such as Transient High Energy Sky and Early Universe Surveyor (THESEUS) \cite{Stratta:2017bwq} are going to play a key role for searches for the time-domain correlations. A detailed study specific to these EM telescopes to study the time domain cross-correlation with the GW data accessible from the current generation GW detectors such as LIGO \cite{TheLIGOScientific:2014jea, Martynov:2016fzi, PhysRevLett.123.231107}, Virgo \cite{Acernese_2014,PhysRevLett.123.231108}, KAGRA \cite{PhysRevD.88.043007,Akutsu:2018axf}, and LIGO-India \cite{Unnikrishnan:2013qwa} will be done in  future work.}

\section{Prospects and conclusion} The novel and unexplored avenue proposed in this work will allow us to study a broad range of scientific applications in the fields of astrophysics, cosmology, and fundamental physics.

\textbf{Astrophysics: }  {\textbf{(i)} This method enables us to search for the EM counterparts of the sub-threshold GW signal and SGWB sources.} \textbf{(ii)} The magnitude and shape of the correlation signal for different characteristic times $\Delta t_{f\nu}$ will explore the energy budget of the astrophysical sources and their temporal evolution up to high redshift. \textbf{(iii)} The time-domain correlation between the sub-threshold (or SGWB) and EM signals will help distinguish  BNS and NS-BH systems from BBHs (or sources without EM counterparts). 
\textbf{(iv)} The inference of the redshift of the source and the accurate sky localization of the GW sources from any EM counterparts is going to help in identifying the host galaxy.  {Hence the population of the GW hosts can be studied up to high redshift. \textbf{(v)} The redshift distribution of the merger rates also  probed by this method will be useful for  distinguishing astrophysical compact objects from primordial black holes in the case  where the latter constitutes most of the dark matter.}

\textbf{Cosmology and Fundamental physics:} \textbf{(i)} In the presence of a characteristic time-scale, the measurement of time-dilation will provide direct evidence of the expansion of the Universe  {and can be used to identify the redshift to the sources contributing to the sub-threshold GW signal and  SGWB using Eq. \eqref{basic-4}}, \textbf{(ii)}  {The observed time delays between the GWs and the EM signals makes it possible to study the speed of propagation of both signals in space-time. The constraints on the differences between the speed of propagation of  GWs $c_{GW}$ and EM waves $c_{EM}$, defined as $\Delta c_{EM-GW}/c_{EM} = c_{EM}\Delta t_{f\nu}/D_l(z)$, should be improved by  orders of magnitude (by a factor $(26\, \text{Mpc}/D_l(z))$) relative to the existing bounds from GW170817 \cite{PhysRevLett.119.161101, Abbott:2017xzu}, due to the large luminosity distance $D_l(z)$ accessible for the sub-threshold GW signal and SGWB sources. For example, sources contributing to the SGWB signal from redshift $z=5$ are going to provide stronger constraints by a factor $\sim 10^{4}$ than the existing bound from GW170817  \cite{PhysRevLett.119.161101, Abbott:2017xzu}. Moreover, all-sky averaging is going to reduce the uncertainties associated with the individual sources and will also improve the constraints by $\sqrt{N_pT_{obs}}$.} \textbf{(iii)} The dispersion of GWs $E^2= p^2c^2 + m^2c^4$ for different frequencies can be tested by comparing the time differences between the propagation of the GW signal and the EM signal $\Delta t_{f\nu}$. \textbf{(iv)} The comparison of the time delays for different frequencies is going to test the theory of gravity from the equivalence of the Shapiro time delay between the GW signal and the EM signal due to gravitational lensing \cite{PhysRevLett.13.789}. Using the sub-threshold GW signal or SGWB signal originating from different redshifts, a tomographic estimation of the Shapiro time delay can be measured. \textbf{(v)} With the availability of all-sky searches for the neutrino background \cite{Aartsen:2020mla}, the temporal cross-correlation between GWs and neutrinos, and between EM signals and neutrinos can be explored. 

The time-domain correlation will open a unique window to identify the EM counterparts to the sub-threshold GW signal and the SGWB using the network of ongoing/upcoming detectors \cite{Aasi:2013wya} such as LIGO \cite{TheLIGOScientific:2014jea, Martynov:2016fzi, PhysRevLett.123.231107}, Virgo \cite{Acernese_2014,PhysRevLett.123.231108}, KAGRA \cite{PhysRevD.88.043007,Akutsu:2018axf}, LIGO-India \cite{Unnikrishnan:2013qwa}, and in the future with LISA \cite{2017arXiv170200786A}, Cosmic Explorer \cite{Reitze:2019iox}, and the Einstein Telescope \cite{Punturo:2010zz}. Our proposed method will exploit the universal nature of the time-lag between the emission of the GW and EM signals (detectable from the ongoing/upcoming missions \cite{2009arXiv0912.0201L, 2012arXiv1208.4012G, 2013arXiv1305.5425S, Aghamousa:2016zmz,2010arXiv1001.0061R, Stratta:2017bwq, Dore:2018kgp, Dore:2018smn, Bellm_2018, Dyer:2020dmo}). This can be used to determine the redshifts of the GW sources contributing to the sub-threshold events and SGWB. Using this method, several aspects of astrophysics, cosmology, and fundamental physics can be explored. The time-domain correlation approach proposed here is also applicable to sub-threshold GW events. This new area of research using time-domain multi-messenger multi-frequency signals potentially develops a new vision for the exploration of the cosmos at high redshift.

\begin{acknowledgements}
The authors thank Sharan Banagiri for carefully reviewing the manuscript and making valuable suggestions during the LSC internal review. SM. would like to thank Joseph Romano for their comments and insightful discussions on the work. SM acknowledges useful comments during the LIGO working group presentation from Thomas Callister, Shivaraj Kandaswamy, Andrew Matas, and Sanjit Mitra.  This analysis was carried out at the Horizon cluster hosted by Institut d'Astrophysique de Paris. We thank Stephane Rouberol for smoothly running the Horizon cluster. SM is supported by the research program Innovational Research Incentives Scheme (Vernieuwingsimpuls), which is financed by the Netherlands Organization for Scientific Research through the NWO VIDI Grant No. 639.042.612-Nissanke. SM  is also supported by the Delta ITP consortium, a program of the Netherlands Organisation for Scientific Research (NWO) that is funded by the Dutch Ministry of Education, Culture, and Science (OCW).  
\end{acknowledgements}
\def\urlprefix{}
\def\url#1{}
\bibliography{main_PRD.bib}
\bibliographystyle{apsrev}
\end{document}